\def\virg#1{``#1"}

\def\eqi{\begin{equation}}
\def\eqf{\end{equation}}
\def\eqia{\begin{eqnarray}}
\def\eqfa{\end{eqnarray}}

\def\ton#1{\left(#1\right)}

\documentclass{aastex}

\usepackage{hyperref}
\usepackage{float}
\usepackage{amsmath,textgreek,w-greek,wasysym}
\usepackage{amscd,lineno}
\usepackage{amssymb}
\usepackage{wasysym}
\usepackage{graphicx,epsfig}
\usepackage{txfonts}
\usepackage{xr-hyper}

\RequirePackage{color}

\newcommand{\emaila}{lorenzo.iorio@libero.it}

\allowdisplaybreaks[1]

\begin{document}
\title{Is the recently proposed Mars-sized perturber at $65-80~\textrm{\textcolor{black}{AU}}$ ruled out by the Cassini ranging data?}

\shortauthors{L. Iorio}

\author{Lorenzo Iorio\altaffilmark{1} }
\affil{Ministero dell'Istruzione, dell'Universit\`{a} e della Ricerca
(M.I.U.R.)-Istruzione
\\ Permanent address for correspondence: Viale Unit\`{a} di Italia 68, 70125, Bari (BA),
Italy}

\email{\emaila}

\begin{abstract}
Recently, the existence of a pointlike pertuber PX with $1~m_{\mars}\lesssim m_\textrm{X}\lesssim 2.4~m_\oplus$ \textcolor{black}{(the symbol \virg{\mars}~denotes Mars)} supposedly moving at $65-80~\textrm{\textcolor{black}{AU}}$ along a moderately inclined orbit has been hypothesized in order to explain certain features of the midplane of the Kuiper Belt Objects (KBOs). \textcolor{black}{We preliminarily selected two possible scenarios for such a PX, and numerically simulated its effect on the Earth-Saturn range $\rho\ton{t}$ by varying some of its orbital parameters over a certain time span; then, we compared our results with some existing actual range residuals.} By assuming $m_\textrm{X} = 1~m_{\mars}$ and a circular orbit\textcolor{black}{, such a} putative new member of our Solar System \textcolor{black}{would nominally perturb $\rho\ton{t}$ by a few km over $\Delta t  = 12~\textrm{yr}~(2004-2016)$. However, the Cassini spaceraft accurately measured $\rho\ton{t}$ to the level of $\sigma_\rho\simeq 100~\textrm{m}$. } Nonetheless, such a scenario should not be considered as \textcolor{black}{necessarily} ruled out since the Cassini data were reduced so far without explicitly modeling any PX. Indeed, a NASA JPL team recently demonstrated that an extra-signature as large as 4 km affecting the Kronian range would be almost completely absorbed in fitting incomplete dynamical models, i.e. without PX itself, to such simulated data, thus not showing up in the standard  post-fit range residuals. \textcolor{black}{Larger} anomalous signatures would \textcolor{black}{instead} occur for $m_\textrm{X} > 1~m_{\mars}$\textcolor{black}{. Their nominal} amplitude could be as large as $50-150~\textrm{km}$ for $m_\textrm{X} =  2.4~m_\oplus$, thus making less plausible their existence.
\end{abstract}

keywords{
gravitation--celestial mechanics--Kuiper belt: general--Oort cloud--space vehicles
}
\section{Introduction}
\textcolor{black}{The possibility that the remote outskirts of the Solar System may hide some still undiscovered object of planetary size has often occurred in the last decades. It has arisen so far in a variety of more or less sound theoretical scenarios generally implying its putative effects on other known bodies; see, e.g.,
\citet{1993AJ....105.2000S,1999Icar..141..354M,1991AJ....101.2274H,
2000MNRAS.318..101C,2003EM&P...92..447M,2004Icar..171..516M,2005AJ....130.1939Z,2008AJ....135.1161L,
2011ApJ...726...33F,2011Icar..211..926M,2014MNRAS.443L..59D,2016A&A...590L...2B,BroBaAJ2016,2016AJ....152..221S,2017AJ....153...63S}. } Recently, \citet{VolkMalhotra017} suggested the existence of a putative new, relatively distant major body of the Solar System whose gravitational tug would allow to explain the deviation of the measured midplane of the Kuiper Belt Objects (KBOs) from the theoretical expectations based only on the currently known planets.
Here, we will follow the convention of \textcolor{black}{generically denoting} the proposed unseen object as X or PX, in which \virg{X} is not meant as the Roman numeral for ten, but simply as \virg{unknown}. If one wanted to, one could even coin provisionally a dedicated name to better distinguish such a scenario \citep{VolkMalhotra017}, characterized by
$m_\textrm{X}\simeq 1~m_{\mars}$ with an upper limit $m_\textrm{X} \lesssim 2.4~m_\oplus$ and $a_\textrm{X} \simeq 65-80~\textrm{\textcolor{black}{AU}}$, from the different one recently proposed by \citet{BaBroAJ2016,BroBaAJ2016} for the so-called Planet Nine, or Telisto \citep{2017Ap&SS.362...11I}, which is a hypothetical more massive and remote new planet with $m_\textrm{X} \simeq 10~m_\oplus,~a_\textrm{X} \simeq 700~\textrm{\textcolor{black}{AU}},~e_\textrm{X} \simeq 0.6,~I_\textrm{X} \simeq  30~\textrm{deg},~\omega_\textrm{X} \simeq 150~\textrm{deg}$. A possible choice\textcolor{black}{,} which captures some of the main features of the putative new member of the Solar System outlined by \citet{VolkMalhotra017}\textcolor{black}{,} could be Taraktor\footnote{From \uptau\upalpha\uprho$\acute{\upalpha}$\upkappa\uptau\upomega\uprho: disturber. It is likely better than the anodyne and uninformative Planet Ten, which is gaining popularity on the Internet at the time of writing. What if Planet Nine/Telisto proposed by \citet{BaBroAJ2016} was not finally found? Would the body put forth by \citet{VolkMalhotra017} become the new Planet Nine? Not to mention even a possible\ldots Planet Eleven in prospect \citep{Eleven017}!}.
In this note, we propose to preliminarily  put to the test its existence as predicted by \citet{VolkMalhotra017} in an independent way  by looking at the dynamical effects exerted by it  on the Earth-Saturn range $\rho\ton{t}$, which is currently known with an accuracy of the order of $\sigma_\rho\simeq 100~\textrm{m}$  thanks to the accurate \textcolor{black}{radio-tracking} data collected over the last 12 yr by the Cassini spacecraft during its exploration of the Kronian\textcolor{black}{\footnote{\textcolor{black}{From K\uprho$\acute{\upo}\upnu\upo\upvarsigma$ (\virg{Cronus/Cronos/Kronos}), corresponding to the Roman deity Saturn.}  }} system \citep{2016DPS....4812007F}.
\textcolor{black}{We will not re-process the Cassini tracking data by including the action of PX. Instead, we will compare the most recent existing post-fit Kronian range residuals\textcolor{black}{\footnote{\textcolor{black}{They are statistically compatible with zero, and look like a somehow unstructured band with semiamplitude of the order of about 100~m. See, e.g., figure 5 of \citet{2014PhRvD..89j2002H}, figure 35 of \citet{2014IPNPR.196C...1F}, figure 1 of \citet{2016DPS....4812007F}, or the blue dots of figure 1 of \citet{2016A&A...587L...8F}.}  }}, obtained without explicitly taking PX into account, with our simulated time series of the range perturbation due to the putative perturber. This allows to infer some plausible, independent insights concerning the model proposed by \citet{VolkMalhotra017} for it.}
\section{The numerically simulated range time series due to PX}
\begin{figure*}
\centerline{
\vbox{
\begin{tabular}{cc}
\epsfxsize= 7.8 cm\epsfbox{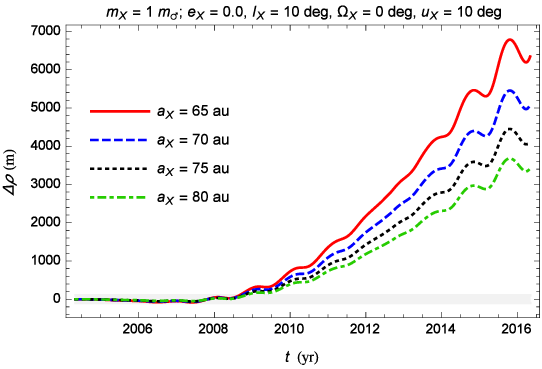} & \epsfxsize= 7.8 cm\epsfbox{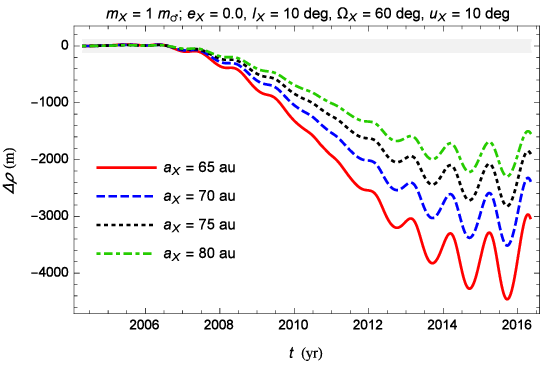}\\
\epsfxsize= 7.8 cm\epsfbox{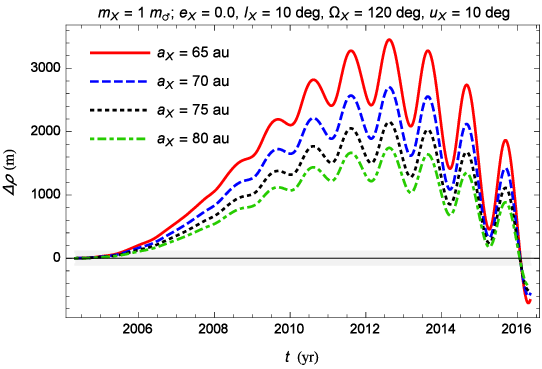} & \epsfxsize= 7.8 cm\epsfbox{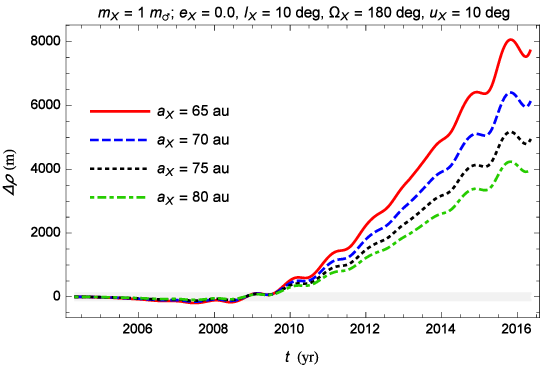}\\
\epsfxsize= 7.8 cm\epsfbox{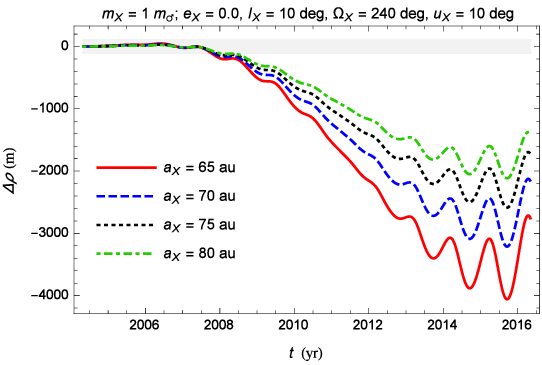} & \epsfxsize= 7.8 cm\epsfbox{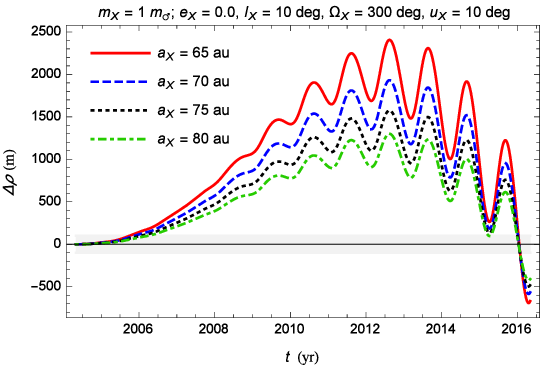}\\
\end{tabular}
}
}
\caption{Numerically simulated nominal Kronian range perturbation $\Delta\rho\ton{t}$, in m,  induced by a Mars-sized perturber PX $\ton{m_\textrm{X}=1 ~m_{\mars}}$ \citep{VolkMalhotra017} over a time span $\Delta t = 12~\textrm{yr}$ long for different values of its semimajor axis $a_\textrm{X}$ and longitude of the ascending node $\Omega_\textrm{X}$. PX is assumed to be located close to the ecliptic near the ascending node $\ton{I_\textrm{X}=10~\textrm{deg},~u_\textrm{X}=10~\textrm{deg}}$ along a mildly inclined, circular orbit  $\ton{e_\textrm{X}=0}$. The gray shaded horizontal band in each picture has a semi-amplitude of $100~\textrm{m}$, and represents the \virg{standard} post-fit range residuals of Saturn produced by processing the Cassini data  without explicitly modeling the dynamics of any PX; cfr. with Figure 1 of \citet{2016DPS....4812007F}.
}\label{figura1}
\end{figure*}

We numerically integrated the equations of motion of all the major bodies of the Solar System in rectangular Cartesian coordinates over the adopted time span $\Delta t = 12~\textrm{yr}$ with and without PX\textcolor{black}{. Both} the integrations shared the same initial conditions for the known planets retrieved from the WEB interface HORIZONS by JPL, NASA. Then, we numerically calculated the corresponding time series $\rho\ton{t}$ for the Earth-Saturn range and took their difference $\Delta\rho\ton{t}$\textcolor{black}{, thus} obtaining the nominal range signature due to PX. Figure \ref{figura1} shows the outcome of our numerical simulations for it as due to a Mars-sized $\ton{m_\textrm{X}=1~m_{\mars}}$ distant $\ton{a_\textrm{X} = 65-80~\textrm{\textcolor{black}{AU}}}$ perturber PX \citep{VolkMalhotra017} moving along a circular $\ton{e_\textrm{X} = 0}$ orbit moderately inclined to the ecliptic $\ton{I_\textrm{X} = 10~\textrm{deg}}$ for various orientations of its orbital plane in space set by the longitude of the ascending node $\Omega_\textrm{X}$; its location along the orbit is assumed to be close to the ascending node $\ton{u_\textrm{X} = 10~\textrm{deg}}$. \textcolor{black}{In this respect}, it must be remarked that no hints are provided by \citet{VolkMalhotra017} about the eccentricity, the orientation of the orbital plane and the position along the orbit of their PX.
It can be noted from Figure \ref{figura1} that the maximum amplitude of the expected signal amounts to a few km, being limited to \textcolor{black}{the range} $\sim 1-8~\textrm{km}$. It turns out that, other things being equal, by displacing PX along its assumed circular orbit, i.e. by varying its argument of latitude $u_\textrm{X}$, the patterns of the resulting time series change, contrary to their amplitudes which remain at the few km level. As far as the inclination $I_\textrm{X}$ of the perturber's orbit is concerned, despite \citet{VolkMalhotra017} assumed it to be small, we explored its full range of variation by finding that it does not notably affect the maximum amplitude of the range perturbations which remains at the $\sim\textrm{km}$ level.
%
%
%
%

In each picture of Figure \ref{figura1}, a grey horizontal band $200~\textrm{m}$ wide is depicted. It summarily represents the width of the actual Kronian post-fit range residuals produced so far by processing the Cassini radiotechnical data without explicitly modeling the dynamics of any distant pointlike perturber PX; cfr. with figure 1 of \citet{2016DPS....4812007F}. Our nominal range perturbation time series $\Delta\rho\ton{t}$ due to PX of Figure \ref{figura1} are not confined within such a relatively narrow band, so that it may be argued at first sight that the scenario for PX adopted here is neatly ruled out by the accurate observations of Cassini. Actually, it would be a precipitate and, to a certain extent, incorrect conclusion\textcolor{black}{\footnote{\textcolor{black}{I am indebted to W.M. Folkner, JPL-NASA, for insightful discussions about such a delicate point.}}}. Indeed, all the currently available \virg{standard} Kronian post-fit range residuals  have been produced so far by using dynamical models which include only the gravitational pull of the known major bodies of the Solar System. Thus, it is quite possible that a non-negligible part of an actually existing PX-driven exotic signature may have been removed in the data reduction procedure, being absorbed in, say, the estimated values of some parameters like the initial state vectors of the other planets.
In other words, the absence of  an anomalous signature like those in our Figure \ref{figura1}
\textcolor{black}{in the Cassini range residuals might just be related to the use of incomplete force models in the fit to the observations.}
Indeed,  figure 3 of \citet{2016DPS....4812007F} shows that fitting usual dynamical models, i.e. without PX, to simulated data including the effect of a perturber\footnote{\citet{2016DPS....4812007F} dealt with the Planet Nine/Telisto scenario \citep{BaBroAJ2016,2017Ap&SS.362...11I} encompassing a much more distant and more massive putative planet with respect to the one considered here.} able to induce a range perturbation on Saturn as large as just $\pm 4~\textrm{km}$ (cfr. with figure 2 of \citet{2016DPS....4812007F}) leads to an almost complete removal from the resulting post-fit residuals. Thus, caution is in order before making hasty conclusions about the inconsistency of the scenario depicted in our Figure \ref{figura1} with the observations; in light of the example discussed in \citet{2016DPS....4812007F}, we tend to consider it as still viable, at least as far as the orbital dynamics of the known major bodies of the Solar System is concerned. It can be shown that, in the same conditions as of Figure \ref{figura1}, the semimajor axis of PX should be as large as $a_\textrm{X}\simeq 200-250~\textrm{\textcolor{black}{AU}}$ in order to have the PX-driven range time series fully within the grey band of the standard post-fit residuals.
On the other hand, the situation is less favorable if larger values of $m_\textrm{X}$ come into play. Indeed, if for $m_\textrm{X} = 1~m_{\mars}$  we are just at the edge of the compatibility with the existing Cassini-based Kronian range residuals, for  $1~m_{\mars}< m_\textrm{X} \lesssim 10-20~m_{\mars}\simeq 1-2 ~m_\oplus$, still encompassed in the scenario proposed by \citet{VolkMalhotra017}, the amplitudes of the corresponding nominal range perturbations may be too large to realistically invoke the previously mentioned mechanism of signal removal in the existing \virg{standard} residuals. Suffice it to say that, for $m_\textrm{X} = 2.4~m_\oplus$, which is the upper limit for the perturber's mass proposed by \citet{VolkMalhotra017}, the nominal range perturbations can be as large as $50-150~\textrm{km}$. In this case, it is more difficult that so huge signatures may have not left some statistically non-zero traces in the currently available post-fit residuals even after the unavoidable partial absorption in the data reduction.

\citet{VolkMalhotra017} did not provide insights about the admissible values of the eccentricity $e_\textrm{X}$ of the perturber's orbit. Here, for the sake of concreteness, we will explore the possibility that it is $0.1\leq e_\textrm{X}\leq 0.5$ for $m_\textrm{X} = 1~m_{\mars}$. If, on the one hand, the amplitude of the nominal Kronian range time series \textcolor{black}{does} not change too much for $e_\textrm{X}=0.1$, on the other hand, the situation is different for $e_\textrm{X} = 0.5$. By discarding a priori orbital positions too close to the perihelion, i.e. $f_\textrm{X}\simeq 0~\textrm{deg}$, because of lacking of direct visual evidence \footnote{A Mars-sized body at just $r_\textrm{X}\simeq 32-40~\textrm{\textcolor{black}{AU}}$ would hardly have escaped a direct detection so far}, some of the remaining portions of the orbit may be questioned  \textcolor{black}{based on} our Kronian dynamics-based analysis.
\begin{figure*}
\centerline{
\vbox{
\begin{tabular}{cc}
\epsfxsize= 7.8 cm\epsfbox{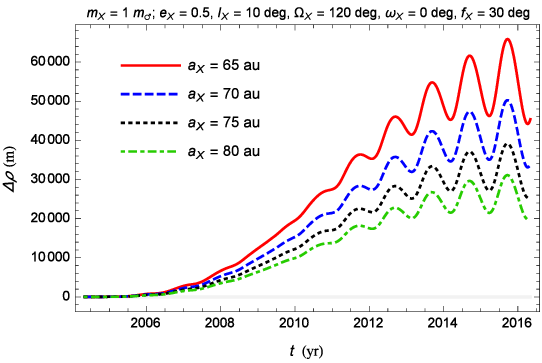} & \epsfxsize= 7.8 cm\epsfbox{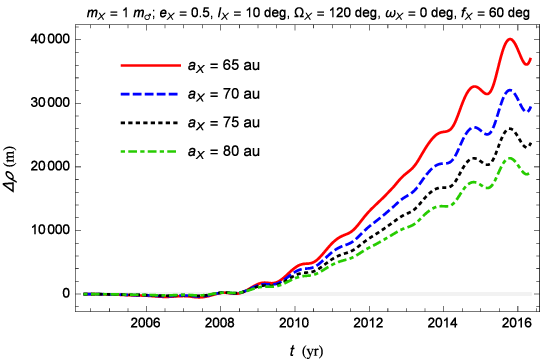}\\
\epsfxsize= 7.8 cm\epsfbox{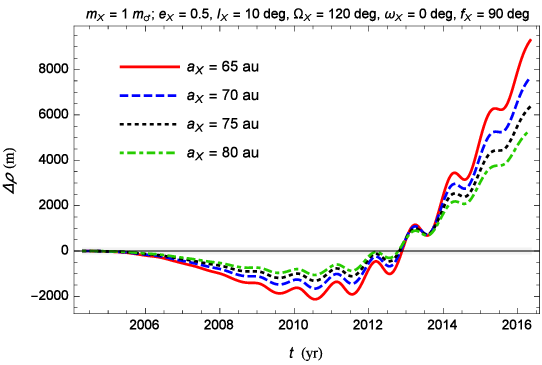} & \epsfxsize= 7.8 cm\epsfbox{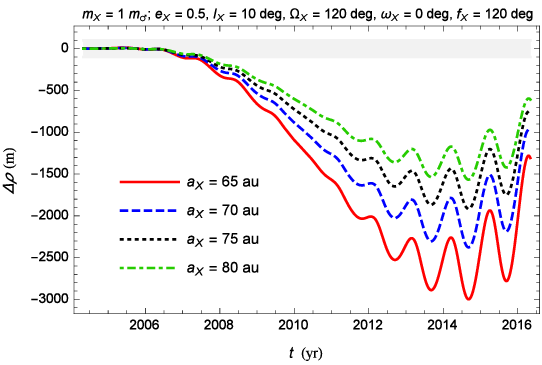}\\
\epsfxsize= 7.8 cm\epsfbox{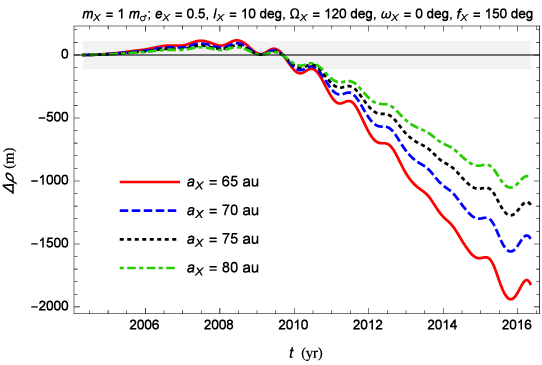} & \epsfxsize= 7.8 cm\epsfbox{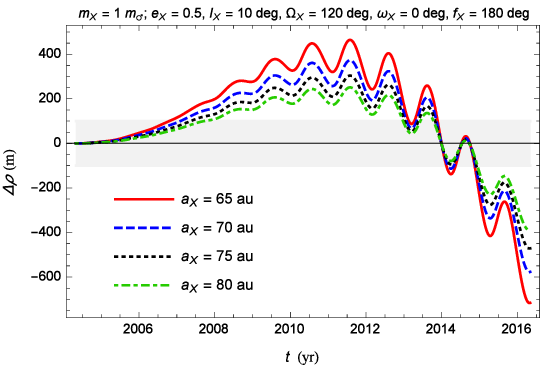}\\
\end{tabular}
}
}
\caption{Numerically simulated nominal Kronian range perturbation $\Delta\rho\ton{t}$, in m,  induced by a Mars-sized perturber PX $\ton{m_\textrm{X}=1 ~m_{\mars}}$ \citep{VolkMalhotra017} over a time span $\Delta t = 12~\textrm{yr}$ long for different values of its semimajor axis $a_\textrm{X}$ and true anomaly $f_\textrm{X}$. PX is assumed to be located far from the perihelion  $\ton{f_\textrm{X}\gg 0~\textrm{deg}}$ along a mildly inclined, eccentric  $\ton{e_\textrm{X}=0.5}$ orbit. The gray shaded horizontal band in each picture has a semi-amplitude of $100~\textrm{m}$, and represents the \virg{standard} post-fit range residuals of Saturn produced by processing the Cassini data  without explicitly modeling the dynamics of any PX; cfr. with Figure 1 of \citet{2016DPS....4812007F}.
}\label{figura2}
\end{figure*}
Indeed, Figure \ref{figura2} shows that the nominal range perturbation of Saturn would be as large as $\simeq 20-60~\textrm{km}$ for $f_\textrm{X}\simeq~30-60~ \textrm{deg}$ for certain values of $\omega_\textrm{X},~\Omega_\textrm{X}$. It must be noted that, however, such constraints are not too surprising since, after all, they would correspond to heliocentric distances as little as $r_\textrm{X}\simeq 34-48~\textrm{\textcolor{black}{AU}}$ which should have allowed a relatively simple direct visual detection of a Mars-sized body. Instead, for orbital positions departing from the perihelion more markedly ($f_\textrm{X}\gtrsim 90~\textrm{deg}$) and approaching the aphelion ($f_\textrm{X}\simeq 180~\textrm{deg}$) it is not possible to rule out dynamically the corresponding anomalous signature because it is at the $\simeq\textrm{km}$ level or even less. Similar results essentially hold also for other values of $\omega_\textrm{X},~\Omega_\textrm{X}$.
Larger values of $e_\textrm{X}$ imply larger portions of the orbit to be dynamically excluded because of a resulting enhancement of the amplitude of the Kronian nominal range perturbations \textcolor{black}{and possible visual detection}.
\section{Summary and conclusions}
The Cassini spacecraft will soon end its life by finally plunging into the atmosphere of the ringed planet on mid of September 2017. The availability of its full ranging data around Saturn, starting in 2004, will offer a unique opportunity to constrain several scenarios  put forth in the latest years involving the presence of a distant, pointlike planetary body PX lurking in the remote peripheries of the Solar System. Latest data reductions, covering the time span $2004-2016$, were performed without explicitly modeling the dynamical action of any hypothetical PX; the resulting standard post-fit Kronian range residuals are statistically compatible with zero, and have an amplitude as little as $\simeq 100$ m. We used them \textcolor{black}{to study} the  PX model recently hypothesized by \citet{VolkMalhotra017} involving a Mars-sized body orbiting at heliocentric distances of about $65-80~\textrm{\textcolor{black}{AU}}$ along a moderately inclined path. \textcolor{black}{We emphasize the preliminary nature of our investigation, which has not pretentions of being an exhaustive study of all possibilities: it detailed just two specific scenarios.} It turned out that, for $m_\textrm{X} = 1~m_{\mars}$ and by assuming a circular orbit for it, such a putative perturber, tentatively dubbed Taraktor (perturber, disturber), would nominally shift the range of Saturn by \textcolor{black}{several} km over 12 yr. This does not necessarily imply that its existence is ruled out by the existing Cassini-based post-fit range residuals of the giant planet since it has been recently demonstrated by a NASA-JPL team that a hypothetical anomalous signal of about $1-4~\textrm{km}$ \textcolor{black}{was almost completely} removed in fitting dynamical models not explicitly encompassing PX to simulated data including its action. Thus, we conclude that, as far as the existing Cassini telemetry is concerned, the PX scenario considered here is likely still compatible with the orbital dynamics of Saturn. On the other hand, according to \citet{VolkMalhotra017}, it is also theoretically possible that the mass of the putative perturber can be as large as up to $2.4~m_\oplus$, implying nominal range perturbations of the order of $50-150~\textrm{km}$. Such hypothetical anomalous signatures might turn out to be too large to have been absorbed \textcolor{black}{by other parameters in}  the data reductions yielding the standard post-fit Kronian range residuals produced so far. Dedicated analyses by reprocessing existing Cassini data \textcolor{black}{would allow for a more rigorous analysis.  Furthermore, it is worth noticing that new missions towards the ice giants \citep{2012ExA....33..753A,2014P&SS..104..122A,2014P&SS..104...93T,NASAgiants017} would be beneficial in order to get better constraints on the mass distribution in the distant regions of the Solar System.}
\bibliography{PXbib,IorioFupeng}{}
%
%
\end{document}